\documentclass[twocolumn,aps,pra,showpacs,preprintnumbers,letterpaper,amssymb,floatfix,superscriptaddress]{revtex4-2}
\usepackage[T1]{fontenc}
\usepackage{natbib}
\usepackage{graphicx}
\usepackage{bm}
\usepackage{color}
\usepackage{amsmath}
\usepackage{xspace}
\usepackage{subfigure}
\usepackage{dcolumn}
\usepackage{textcomp}
\usepackage{longtable}     
\usepackage{url}           
\usepackage[colorlinks,urlcolor=blue,citecolor=blue,linkcolor=blue]{hyperref}
\usepackage{bm}            
\usepackage{appendix}
\usepackage{hyperref}
\usepackage{soul}
\usepackage[normalem]{ulem}
\usepackage{physics}
\usepackage{tabularx}
\usepackage{array}
\usepackage{multibib}

\newcites{supp}{Supplementary References}

\def\NaRb{$^{23}$Na$^{87}$Rb\xspace}

\usepackage{xcolor}

\begin{document}


\title{Creation of ultracold heteronuclear $p$-wave Feshbach molecules}

\author{Fan Jia}
\email{fan.jia@mpq.mpg.de}
\thanks{Current address: Max Planck Institute of Quantum Optics, Garching, Germany}
\affiliation{Department of Physics, The Chinese University of Hong Kong, Hong Kong, China}
\author{Zhichao Guo}
\thanks{Current address: Department of Applied Physics and Science Education,
Eindhoven University of Technology, P. O. Box 513, 5600 MB Eindhoven, The Netherlands}
\affiliation{Department of Physics, The Chinese University of Hong Kong, Hong Kong, China}
\author{Zerong Huang}
\affiliation{Department of Physics, The Chinese University of Hong Kong, Hong Kong, China}
\affiliation{State Key Laboratory of Quantum Information Technologies and Materials,\\
The Chinese University of Hong Kong, Hong Kong SAR, China}
\author{Dajun Wang}
\email{djwang@cuhk.edu.hk}
\affiliation{Department of Physics, The Chinese University of Hong Kong, Hong Kong, China}
\affiliation{State Key Laboratory of Quantum Information Technologies and Materials,\\
The Chinese University of Hong Kong, Hong Kong SAR, China}




\date{\today}

\begin{abstract}
We report the first creation of a bulk sample of ultracold heteronuclear $p$-wave Feshbach molecules in an optically trapped Bose-Bose mixture of $^{23}$Na and $^{87}$Rb atoms. Using loss spectroscopy and binding energy measurements, we systematically characterize the interspecies $p$-wave Feshbach resonances near 284 G. Leveraging this understanding, we use magneto-association to form $p$-wave NaRb Feshbach molecules, producing both pure samples and mixtures of molecules in different $p$-wave orbitals. We further measure the molecular lifetime and identify atom-molecule and molecule-molecule collisions as the dominant loss mechanisms. This work establishes a previously unavailable ultracold molecule platform that combines orbital anisotropy with heteronuclear constituents,  represents a significant step toward realizing tunable $p$-wave interactions in Bose-Bose mixtures, and provides a foundation for exploring non-zero angular momentum molecules.
\end{abstract}

\maketitle

Magnetically tunable Feshbach resonances (FRs) provide a powerful route to controlling interactions in ultracold gases and to creating weakly bound Feshbach molecules (FMs) through magneto-association (MA)~\cite{Chin2010a,Kohler2006}. While most experiments have focused on the familiar $s$-wave regime, higher partial wave resonances are of particular interest because they introduce orbital structure and anisotropic interactions~\cite{Regal2003,Ticknor2004,Zhang2004,Gunter2005,Gaebler2007,Inada2008,Dong2016,Cui2017,zhang2021transition,zhang2023many}. In particular, $p$-wave interactions have enabled studies of orbital pairing and strongly interacting fermions beyond the $s$-wave paradigm~\cite{Gurarie2005,Cheng2005,Levinsen2007,Yu2015,He2016,He2017,He2021,Luciuk2016,Venu2023,Jackson2023}. However, experimental access to ultracold molecules with non-zero orbital angular momentum remains comparatively limited.

A particularly important missing platform is a bulk heteronuclear gas of $p$-wave FMs. Compared with earlier $p$-wave FMs created in homonuclear fermionic systems~\cite{Gaebler2007,Inada2008}, the heteronuclear case combines orbital anisotropy with distinguishable constituents, mass imbalance, and a natural connection to the broader field of ultracold polar molecules. In Bose-Bose mixtures, where all partial waves are allowed between non-identical particles, an interspecies $p$-wave FR coexists with background $s$-wave interactions and has been predicted to support rich few- and many-body behavior, including finite-momentum superfluidity and modified miscibility~\cite{Radzihovsky2009,Choi2011,Deng2024}. Heteronuclear $p$-wave molecules are also appealing as a starting point for studies of anisotropic atom-molecule and molecule-molecule collisions, species-resolved dissociation dynamics~\cite{Greiner2005,Gaebler2007}, and direct access to rotationally excited states of ground-state molecules for DC field induced loss suppression~\cite{Matsuda2020,Li2021,Walraven2024,Lin2026Fermi}, without requiring additional microwave coupling. Despite these prospects, previous bulk-gas experiments had only identified interspecies $p$-wave resonances and molecular binding energies without producing heteronuclear $p$-wave molecular samples~\cite{Papp2007,Dong2016}, while recent heteronuclear $p$-wave molecule production was limited to single molecules in optical tweezers~\cite{Zhang2020}.

The absence of earlier bulk heteronuclear $p$-wave molecular samples likely reflects the stringent conditions required for their production. In a heteronuclear mixture, one must work at low enough temperature and with precise enough magnetic field control to resolve the narrow, split $p$-wave resonance structure, while maintaining sufficient interspecies overlap and phase-space density for detectable association. Moreover, the comparatively weak coupling between the open and closed channels of $p$-wave resonances can make MA inefficient, requiring slow magnetic field ramps while inelastic atom-molecule and molecule-molecule collisions can strongly limit the formation and survival of the FMs. The present work is enabled by improved control over these competing requirements.

\begin{figure*}[t]
\begin{center}
\includegraphics[width=0.8\linewidth]{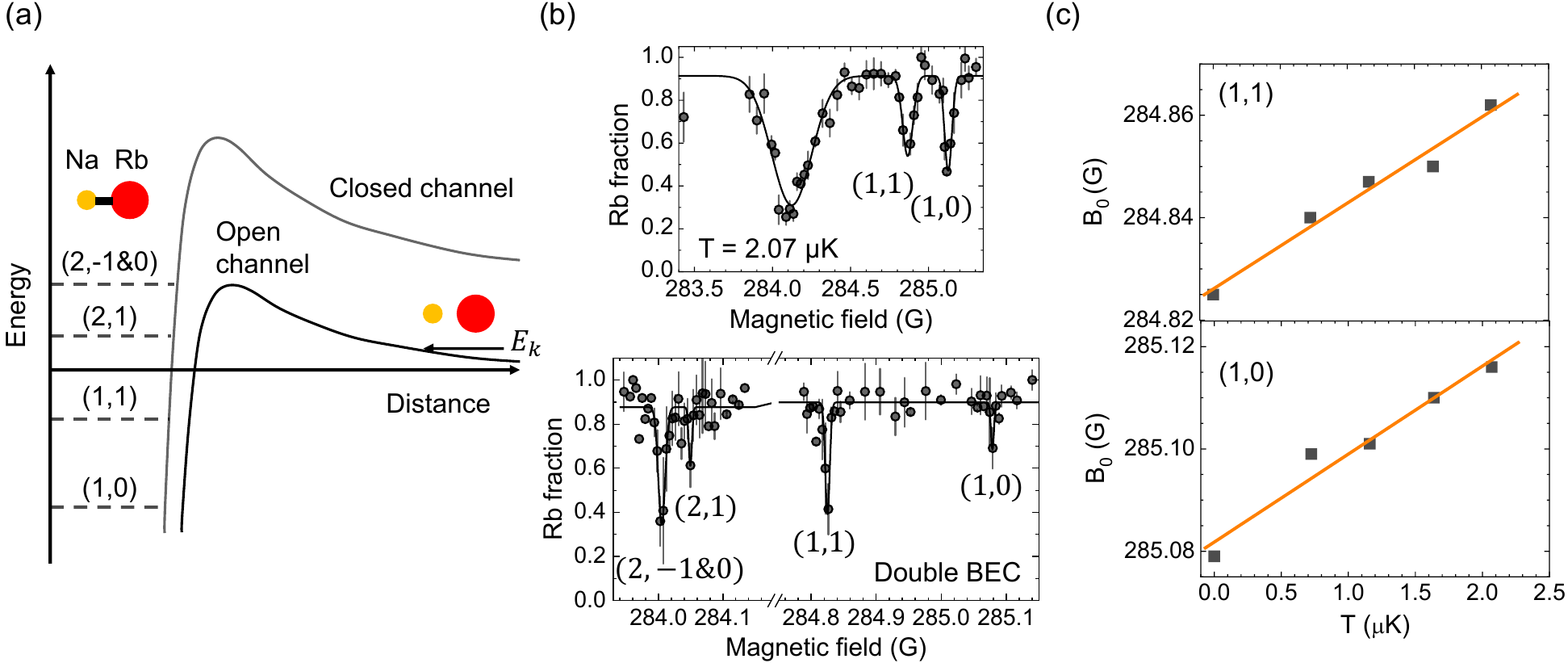}
\end{center}
\caption{Probing $p$-wave Feshbach resonances with loss spectroscopy. (a) Two-channel schematic of $p$-wave FRs in the heteronuclear Na-Rb mixture. Both bound and quasi-bound states can be supported in the closed channel. (b) (top) Remaining fractional Rb atom numbers as a function of magnetic field after a 50 ms holding time for a thermal mixture at 2.07 $\mu$K and (bottom) the same measurement for a double BEC sample. For the two peaks of the $m_f = 2$ resonance, the holding time is 25 ms, while for the two peaks of the $m_f = 1$ resonance, a longer holding time of 120 ms is used. The faster losses observed for the $m_f = 2$ resonance in both indicate that the coupling strength of this resonance is stronger than that of the $m_f = 1$ resonance. (c) Loss peaks $B_0$ as a function of sample temperature $T$ for the $(m_f = 1, m_l = 1)$ (top) and $(1,0)$ (bottom) peaks. The separation between the two $m_f = 2$ resonances is approximately 40 mG and they overlap at temperatures above 700 nK. The orange lines are linear fits to the data.}
\label{fig1}
\end{figure*}

We investigate two closely spaced $p$-wave FRs in the $^{23}$Na-$^{87}$Rb Bose-Bose mixture. Based on this characterization, we study the MA process and create bulk samples of $p$-wave NaRb FMs in either pure or mixed $m_l$ states. Here, $m_l$ represents the projection of the relative rotational angular momentum or partial wave $l$, which can be $\pm 1$ or 0 for $p$-wave scattering with $l = 1$. We also find that pure $p$-wave NaRb molecular samples can live long enough for collision studies and for transfer to the molecular ground state. Our results provide an opportunity to study loss behavior of heteronuclear $p$-wave molecules, especially in lower dimensions~\cite{Pricoupenko2008,Peng2014,Zhou2017}, and offer a platform for exploring few-body and many-body physics with high-partial-wave interactions.

The two $p$-wave FRs used in this study were first observed in our previous work using ultracold thermal Na and Rb mixtures in their $\ket{F=1,m_F=1}$ hyperfine states, with a typical sample temperature $T$ of about $1.8~\mu\mathrm{K}$~\cite{Wang2013}. Based on coupled-channel calculations~\cite{Wang2013,Hutson2019}, the two features near $284~\mathrm{G}$ are assigned to the $m_f=2$ manifold as $(m_f,m_l)=(2,-1)$, $(2,0)$, and $(2,1)$, with the $m_l=-1$ and $m_l=0$ components nearly degenerate. Here, $m_f$ is the projection of the total hyperfine angular momentum in the closed channel. The other two features near $285~\mathrm{G}$ are assigned to the $m_f=1$ manifold as $(1,1)$ and $(1,0)$. Fig.~\ref{fig1}(a) shows the two-channel schematic and illustrates relative molecular energies associated with these FRs.

We first characterize the FRs with loss spectroscopy at different sample temperatures. Following the standard procedure detailed in Ref.~\cite{Wang2013}, we prepare an atomic mixture of Na and Rb in a crossed $1070~\mathrm{nm}$ optical dipole trap (ODT). The sample temperature is controlled by varying the final trap depth during evaporative cooling. The FRs are identified by measuring the fractional loss of both Na and Rb atoms after ramping the magnetic field to a target value and holding for a carefully chosen duration.

As shown in the two example measurements in Fig.~\ref{fig1}(b) for $T = 2.07~\mu$K  and for a quasi-pure double BEC, the atom loss depends strongly on temperature. The apparent narrowing of the linewidth for the double BEC compared with the thermal sample is characteristic of non-$s$-wave collisions due to the centrifugal barrier~\cite{SM}. According to our coupled-channel modeling~\cite{Hutson2019,Guo2022}, both FRs exhibit coupling strengths between the open and closed channels that are strong enough to categorize them as broad resonances~\cite{Dong2016,Cui2017}, with the $m_f = 2$ resonance displaying significantly stronger coupling than the $m_f = 1$ resonance, consistent with our observations. However, with the double BEC sample, our measurement shows that the resonances are still narrow in terms of magnetic field.

For the measurement taken with double BEC samples, the four loss peaks, $(2,-1\, \& \,0)$, $(2, 1)$ for the $m_f = 2$ manifold on the left, and $(1, 1)$, $(1, 0)$ for the $m_f = 1$ manifold on the right, are fully resolved. From a double Gaussian fit to the two $m_f = 2$ peaks, we find that the separation between them is about 40 mG. We also observe that for temperatures higher than 700 nK, the two loss peaks start to merge together. At around 2 $\mu$K, they can not be resolved individually, as shown in the measurement when $T = 2.07~\mu$K. On the other hand, the two peaks in the $m_f = 1$ manifold are always well separated.

Besides the varying of the loss widths, the temperature effect also causes shift of the loss peak positions. The positions associated with the $(1, 1)$ and $(1, 0)$ features as a function of $T$ are summarized in Fig.~\ref{fig1}(c). The linear dependence on $T$ is consistent with the theoretical results based on the so-called trimer model (see the SM~\cite{SM} for details).

To further characterize the FRs, we measure the binding energy $E_b$ of the FMs with the magnetic field modulation method (often termed wiggle spectroscopy)~\cite{Thompson2005}. Near the $p$-wave resonance, a pair of Na and Rb atoms will form a FM when the magnetic field modulation frequency $\nu$ is tuned to $|E_b - E_k|/h$, where $E_k$ is the relative kinetic energy of the atom pair. These FMs are highly prone to loss and heating from atom-molecule and molecule-molecule inelastic collisions. Consequently, the resonance can be inferred by monitoring the fractional loss of atoms or changes in the size of the remaining atomic cloud. Due to the thermal Boltzmann distribution at finite temperatures, both the linewidth and the peak position depend on temperature, and the loss and heating spectra are also asymmetric.

\begin{figure}[ht]
\begin{center}
\includegraphics[width = 0.85\linewidth]{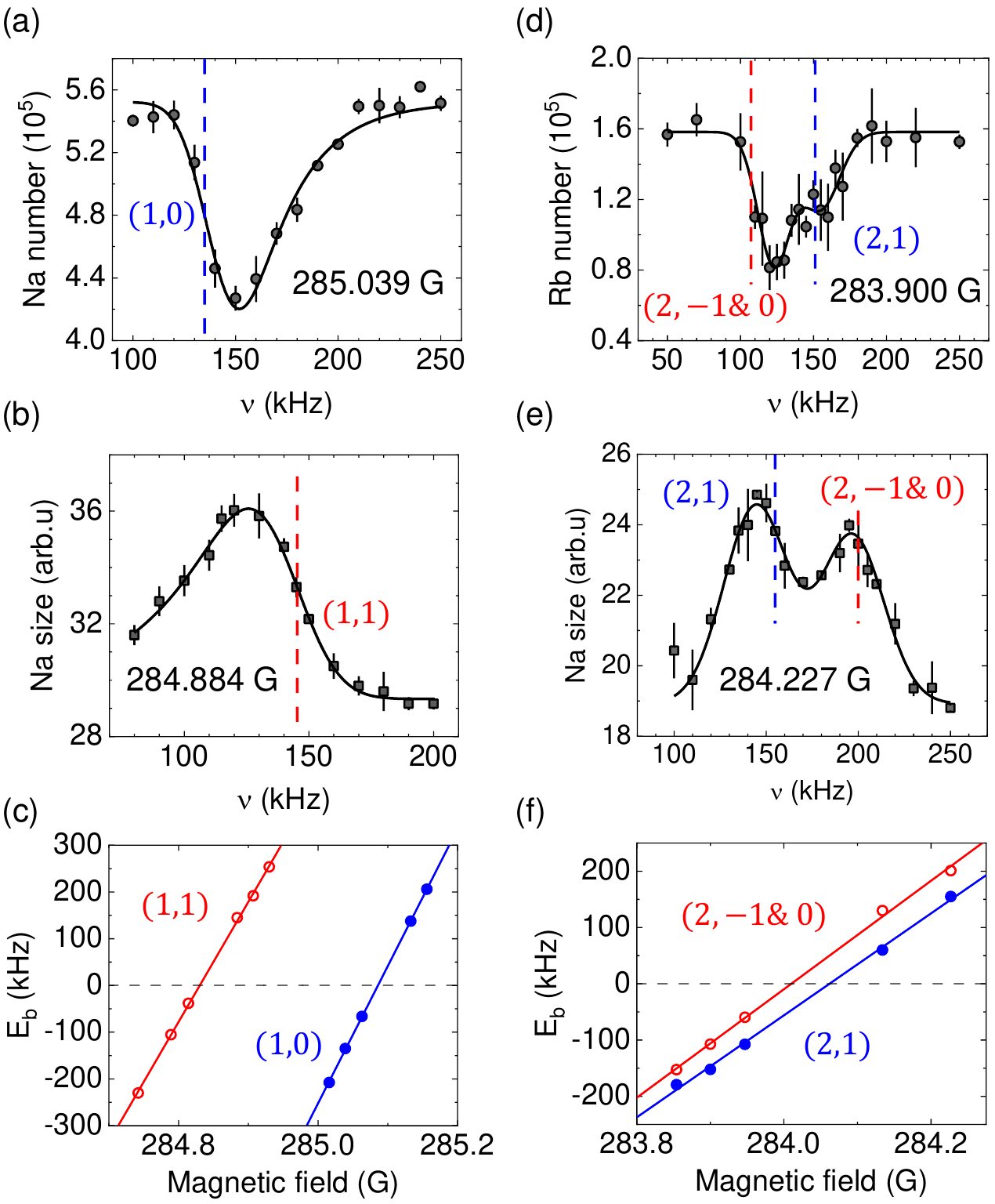}
\end{center}
\caption{Binding energy measurement by magnetic field modulation. (a) Atom loss versus the modulation frequency $\nu$ at 285.039 G for probing the $(1,0)$ resonance. (b) Cloud size versus $\nu$ at 284.884 G for probing the quasi-bound states associated with the $(1,1)$ resonance. (c) The measured binding energy $E_b$ for $(1,1)$ (red circles) and $(1,0)$ (blue circles) resonances. (d), (e) are similar loss and heating measurements for the $m_f = 2$ manifold, respectively. The $E_b$ difference between the two resonances is small but still resolvable. (f) shows the measured $E_b$ for $(2,-1\&0)$ (red circles) and $(2,1)$ (blue circles) resonances. The solid curves in (a), (b), (d), and (e) are fits to Gaussian convoluted with Boltzmann function for extracting $E_b$ (marked by the vertical dashed lines).  The solid lines in (c) and (f) are linear fits to determine the magnetic dipole moments of the FMs relative to the atoms (the results are summarized in Table \ref{tab:p-wave}).}  
\label{fig2}
\end{figure}

An example modulation association spectrum measured with loss of Na atoms for the $(1, 0)$ resonance is presented in Fig.~\ref{fig2}(a). The temperature of the starting atomic mixture for this measurement is around 600 nK. To extract $E_b$, we fit the curve with a phenomenological Gaussian convoluted with a Boltzmann function. As shown in the figure, the molecular binding energy is located on the left side of the loss maximum, consistent with the intuitive expectation that additional energy is required to couple energetic atoms into molecules.

For the current system, the height of the $p$-wave barrier is 388.7 $\mu$K~\cite{Wang2013,Guo2022}. This is enough to support quasi-bound FMs which are subject to fast dissociative decay via tunneling. As the trap depths for both atoms are larger than the binding energy, the fragmented atoms stay in the trap and result in heating. For instance, as shown in Fig.~\ref{fig2}(b), quasi-bound molecules of the $(1, 1)$ resonance are associated at 284.884 G, and $E_b$ is obtained by fitting the change in the Na cloud size after 4 ms time-of-flight (TOF) expansion~\cite{SM}. Notably, in contrast to Fig.~\ref{fig2}(a), the fitted $E_b$ is located on the right side of the peak, indicating that less energy is required to couple energetic atom pairs into quasi-bound molecules.

As shown in the example data in Fig.~\ref{fig2}(d) and Fig.~\ref{fig2}(e), similar loss and heating measurements can also be performed for the bound (283.900 G) and quasi-bound (284.227 G) states of the $m_f = 2$ manifold. As the $(2,-1\&0)$ and $(2,1)$ resonances are very close to each other, the $E_b$ difference between them is also very small. Thanks to the high resolution of the magnetic field modulation technique, the two resonances can still be clearly resolved.    

The measured $E_b$ for the $m_f = 1$ and $m_f = 2$ manifolds are summarized in Fig.~\ref{fig2}(c) and (f). The results show that $E_b$ varies linearly with the magnetic field, indicating a constant fraction of the closed channel in the dimer across the scanned range~\cite{Fuchs2008}. The slope of the linear fit, $\delta \mu_b^{exp}$, corresponds to the difference in magnetic moments between molecule and atom pair. As summarized in Table~\ref{tab:p-wave}, the experimentally measured $\delta \mu_b^{exp}$ agree well with the theoretical values $\delta \mu_b^{th}$ from coupled-channel calculations~\cite{SM}.

\begin{table}[h]
\caption{Experimentally measured $\delta \mu_b^{exp}$ and theoretically predicted $\delta \mu_b^{th}$ values associated with the $B = 284$ G $p$-wave resonances.}
\label{tab:p-wave}
\centering
\begin{tabular}{lcc}
    \hline\hline
    & $\delta \mu_b^{\text{exp}}$ (kHz/G) & $\delta \mu_b^{\text{th}}$ (kHz/G) \\ 
    \hline
    $m_f = 2$, $m_l = -1\&0$ & 980(5) & 943(9) and 980(10)\\ 
    $m_f = 2$, $m_l = 1$  & 914(10) & 900(12) \\ 
    $m_f = 1$, $m_l = 1$ & 2628(7) & 2599(45) \\ 
    $m_f = 1$, $m_l = 0$  & 2887(107) & 2994(30) \\ 
    \hline\hline
\end{tabular}
\end{table}

To create $p$-wave FMs, we use MA by sweeping the magnetic field across a resonance from above. At first glance, the two well-separated peaks of the $m_f = 1$ manifold appear to be the most suitable choices for creating $p$-wave FMs. However, we were not able to observe any signal of FMs using this manifold, despite tuning of the MA parameters over a large range. We attribute this to the weak coupling strength between the open and closed channels for this manifold, which requires a very slow magnetic field sweep rate for efficient FM conversion~\cite{Hobdy2005,Cumby2013}. At the same time, severe atom-molecule inelastic collisions, which likely occur during the sweep, make it experimentally challenging to form detectable FMs with this resonance. On the other hand, the stronger coupling strength of the $m_f = 2$ manifold makes MA with this resonance more feasible, even though the splitting between the two peaks is only 40 mG.

To detect the first signature of $p$-wave FMs, we first ramp the magnetic field to 285.3 G, which is above both manifolds. We then sweep the magnetic field to 283.90 G, which is below both manifolds, at a constant rate for MA. Subsequently, we quench the magnetic field to 279.23 G, turn off the ODT, and apply a 160 G/cm gradient pulse for 1.5 ms to separate the FMs from the remaining atoms. The magnetic field quench increases the binding energy which are important for preventing accidental dissociation and separating FMs from atoms. As shown in Table~\ref{tab:p-wave}, the differential magnetic moment $\delta \mu_b$ is moderate but enough for us to separate the molecules from the remaining atoms, especially the lighter Na atoms, using the gradient pulse~\cite{Herbig2003}.

For detection, after the gradient pulse, we dissociate the FMs using magnetodissociation (MD) by ramping the magnetic field reversely to above the resonance and then detect the fragmented Na atoms via the high-field absorption imaging method~\cite{Jia2020}. The high-field imaging protocol involves two laser beams: an optical pumping beam that pumps the atoms to the $\ket{2,2}$ state, and a $\sigma^+$-polarized probe light on the cycling transition. This method can also directly probe FMs without MD, as the optical pumping beam can dissociate the FMs via photodissociation (PD). A major difference between the two dissociation methods is that PD typically results in much greater heating, especially for the lighter Na atoms. Additionally, MD can selectively dissociate FMs of different $m_l$ states, while PD lacks such selectivity because the excited-state linewidth is much larger than the binding energies of all the FMs.

\begin{figure}[t]
\begin{center}
\includegraphics[width = 0.85\linewidth]{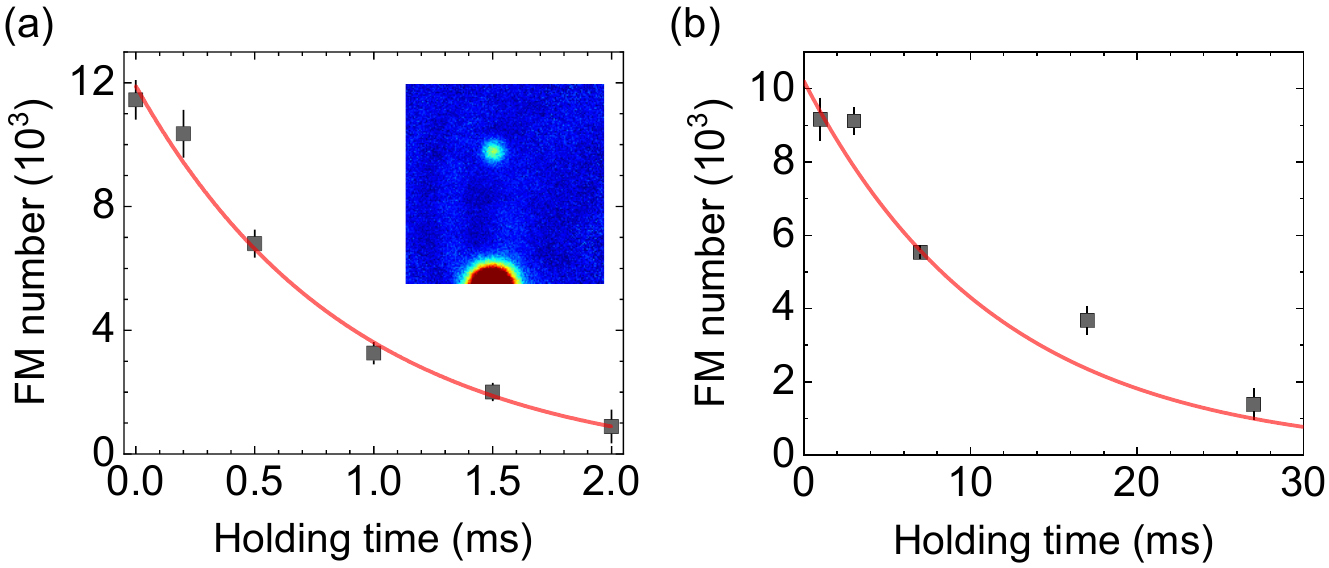}
\end{center}
\caption{First signals of $p$-wave NaRb FMs created via MA. (a) The image in the inset shows the signature of FMs (top small cloud) observed after separating them from residual atoms (bottom cloud) with a magnetic field gradient. The lifetime of the FMs in the presence of residual atoms is only 0.8(1) ms due to fast atom-molecule collisions. (b) After removal of the residual atoms, the lifetime of the pure FMs is extended to 12(3) ms. The FMs, created by sweeping the magnetic field across all resonances, are distributed among $(2,-1\&0)$ and $(2,1)$ states. The red solid curves are exponential fits for extracting the lifetimes. } 
\label{fig3}
\end{figure}

The inset of Fig.~\ref{fig3}(a) shows the first signal of heteronuclear $p$-wave FMs obtained via MD by ramping the magnetic field to 285.3 G, which dissociates FMs of all $m_l$ states. The FMs appear as an additional cloud above the much larger residual Na atom cloud. At an optimized MA magnetic field sweeping rate of 0.7 G/ms, up to $1.2 \times 10^4$ FMs are produced from an initial sample of $4 \times 10^5$ Na and $2 \times 10^5$ Rb atoms. This corresponds to a 6\% conversion efficiency of Rb atoms into molecules, comparable to that previously achieved using the 347 G $s$-wave resonance~\cite{Wang_2015}.

By varying the coexistence time of the molecules and the remaining atoms in the trap at 279.23 G, we observe fast loss of FMs caused by atom-molecule collisions. Because the number of atoms far exceeds the number of molecules, the former can be deemed constant, and a one-body loss model can be applied to fit the molecular loss. As shown in Fig.~\ref{fig3}(a), an exponential decay fit to the data yields a short FM lifetime of 0.8(1) ms.

To create a sample of FMs without residual atoms, we apply a sequence of microwave and blast light pulses after the MA process to remove the remaining Na and Rb atoms~\cite{SM}. At the end of the atom removal procedure, we typically obtain $10^4$ FMs. As the magnetic field sweep crosses the whole $m_f = 2$ manifold, the FMs form as a mixture of $(2,-1\&0)$ and $(2,1)$ states. As presented in Fig.~\ref{fig3}(b), the lifetime of the pure FMs is 12(3) ms, much longer than that of the atom-molecule mixture.

\begin{figure}[htb]
\begin{center}
\includegraphics[width = 0.9\linewidth]{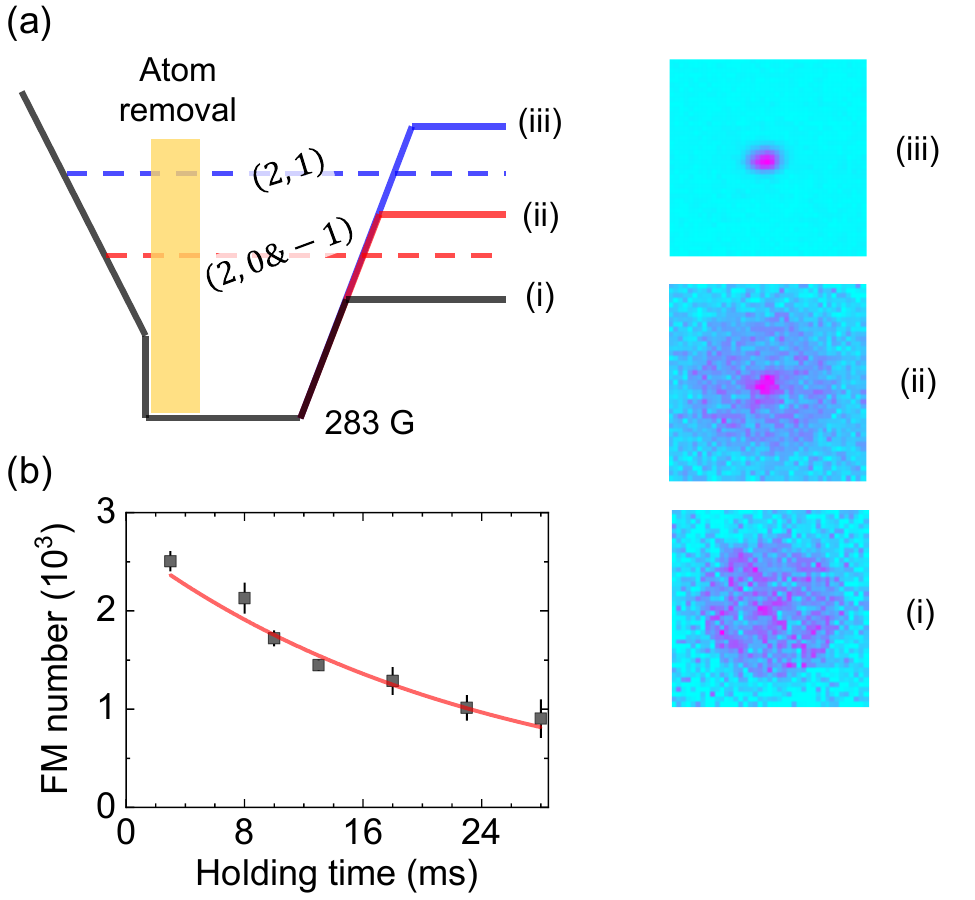}
\end{center}
\caption{Creation of pure $(2, 1)$ FMs. (a) $(2, -1\&0)$ and $(2, 1)$ FMs can be distinguished by the significantly different amount of heating of PD and MD during high-field imaging. The red and blue dashed lines represent the positions of the $(2, -1\&0)$ and $(2, 1)$ resonances, respectively. The state of the FMs, as shown by the absorption images on the right, is determined by the $B$ field endpoint of the MA process (see text for details). The quench following MA is sufficiently rapid to prevent additional association. (b) The exponential fitted lifetime of the pure $(2, 1)$ FMs is 23.5(1) ms and is likely limited by excitation from the trap laser. The endpoint of the MA process is 284.056 G, which lies between the two resonances within region (ii).  
}
\label{fig4}
\end{figure}

The different $m_l$ FMs can be distinguished by leveraging the distinct state selectivity and heating effects of MD and PD. As illustrated in Fig.~\ref{fig4}(a), when the magnetic field for MD is stopped below the $(2, -1\&0)$ resonance [region (i)], all FMs are detected via PD during high-field absorption imaging. To increase the kinetic energy of the fragmented atoms from PD, the optical pumping/PD light is tuned 150 MHz above the dissociation threshold. The near 5.4 mK energy imparted to the fragmented Na atoms leads to a large and dilute cloud due to the rapid expansion during the 50 $\mu$s time interval between PD and the probe pulse. On the other hand, when the magnetic field is ramped between the two resonances [region (ii)], the $(2, -1\&0)$ FMs are already dissociated by MD. As the heating from MD is much smaller and the fragmented atoms are not significantly heated by the imaging light pulses, this portion of the FMs appears as a more compact region in the image center. In contrast, the $(2, 1)$ FMs are dissociated by PD and appear as a much larger and more dilute region. Finally, when the magnetic field is ramped above the $(2, 1)$ resonance [region (iii)], all FMs are dissociated by MD, and only a much smaller cloud is observed. With this detection method, we also reconfirm that no $m_f = 1$ FMs are created, despite crossing the $m_f = 1$ peaks during the MA process.

The capability to distinguish different $m_l$ states also provides a reference for creating pure $(2, 1)$ FMs. To this end, during MA, we sweep the magnetic field ending point to region (ii). The magnetic field is then abruptly quenched to 283 G at a rate too fast to create any additional FMs. After removing the residual atoms, we obtain a pure sample of FMs at 283 G. We then ramp the magnetic field back to region (ii) for detection and to distinguish the $m_l$ occupation of the FMs.

After fine adjustment of the MA ending point, we find that sweeping the magnetic field to 284.056 G reliably eliminates the compact center, leaving only a dilute cloud in the absorption images. This provides evidence that no $(2, -1\&0)$ FMs are created and that all the signals are from $(2, 1)$ FMs. After confirming this, we can then detect the pure $(2, 1)$ sample using MD by ramping the magnetic field to region (iii) for an improved SNR. Following this procedure, we can routinely produce a pure $(2, 1)$ sample containing approximately $3\times 10^3$ FMs, with a typical peak density of $7\times 10^{10}\,\rm{cm^{-3}} $ and a typical temperature of 2.8 $\mu$K. It should be noted, however, that the in-trap sample temperature is likely lower, since the MD ramp during the measurement can heat up the resultant atoms.

As shown in Fig.~\ref{fig4}(b), the trap lifetime of the pure $(2, 1)$ FM sample is 23.5(1) ms. We believe that the current lifetime is affected by excitation from the trapping light, which is provided by a multi-mode laser. Previous studies of $s$-wave FMs have shown that the lifetime can be significantly increased by switching to a single-frequency laser~\cite{Wang_2015,Wang2019,Guo2016}. Nevertheless, even the current lifetime is sufficient for Raman transfer to the ground state.

To summarize, we have successfully created samples of heteronuclear $p$-wave FMs using an ultracold mixture of Na and Rb atoms. We further characterized their lifetime and identified atom-molecule and molecule-molecule collisions as the dominant loss mechanisms. Although the molecular lifetime remains likely limited by the excitation from the trapping light, future upgrades to the experimental setup should enable detailed studies of collisional dynamics in this heteronuclear $p$-wave molecular system. These advancements may also allow us to explore theoretical predictions that, in lower dimensions, the inelastic collision rate of $p$-wave FMs may be significantly reduced~\cite{Zhou2017}, paving the way for the realization of long-lived $p$-wave FMs.

The photodissociation of heteronuclear $p$-wave Feshbach molecules also provides an opportunity for species-resolved studies of the resulting Na and Rb atoms, including their momentum and angular distributions, which reflect the non-zero orbital angular momentum of the parent molecules~\cite{Gaebler2007}. In the present setup, detection perpendicular to the quantization axis limits our ability to resolve these angular features and correlations. Future improvements, such as coincidence-sensitive detection and probing along the quantization axis, could enable more detailed studies of dissociation dynamics and fragment correlations~\cite{Greiner2005}.

\begin{acknowledgments}
We are grateful to Bin Zhu, Yue Cui, Zhendong Zhang, Mingyang Liu and Shizhong Zhang for valuable discussions, as well as Xinyuan Gao and Yangqian Yan for their careful reading of our manuscript. This work is supported by Quantum Science and Technology--National Science and Technology Major Project of China (2024ZD0300600), the Hong Kong RGC General Research Fund (Grants 14304323 and 14302722) and the Collaborative Research Fund (Grant No. C4050-23G), and Guangdong Provincial Quantum Science Strategic Initiative (Grant No. GDZX2303002).
\end{acknowledgments}

\bibliography{P-wave}

\renewcommand\thefigure{\thesection S\arabic{figure}} 
\renewcommand\theequation{\thesection S\arabic{equation}} 
\setcounter{figure}{0}
\setcounter{equation}{0}

\section*{Supplemental Material}

\section{Section S1: TEMPERATURE-DEPENDENT ATOM-LOSS SPECTRA}

A coupled-channel ``trimer'' model was proposed to account for the temperature-dependent three-body recombination resonance~\cite{Maier2015}. In this model, atoms from the scattering asymptote in the entrance channel can temporarily form trimers in closed channels. These trimers can subsequently break up into a weakly bound dimer and an atom in one of the open channels, followed by energy release and atom loss. The recombination rate coefficient at collision energy $E_3$ for this process is given by:
\begin{equation}
L_3(E_3, B) = (2 \lambda + 1) \frac{192 \pi^2}{k_3^5} \frac{\hbar k_3}{\mu_3} \left| S(E_3, B) \right|^2,
\label{eq1}
\end{equation}
where $k_3$ is the relative wave vector, $\mu_3 = \sqrt{m_1 m_2 m_3 / (m_1 + m_2 + m_3)}$ is the three-body reduced mass, and $\lambda$ represents the relative angular momentum. The $S$-matrix element $\left| S(E_3, B) \right|^2$ can be expressed as:
\begin{equation}
\left| S(E_3, B) \right|^2 = \frac{\Gamma(E_3) \Gamma_{\mathrm{br}}}{[E_3 - \mu (B - B_0)]^2 + [\Gamma_{\mathrm{tot}}(E_3) / 2]^2},
\label{eq2}
\end{equation}
where $B_0$ is the trimer resonance location and $\mu$ is the relative magnetic moment between the trimer and the entrance channel. $\Gamma_{\mathrm{tot}}(E_3)$ is the total energy width, which is the sum of the entrance-channel energy width $\Gamma(E_3) = A_\lambda E_3^{\lambda + 2}$ and the decay rate of the resonance into the dimer and atom, $\Gamma_{\mathrm{br}}$.

This trimer model effectively explains the experimentally observed temperature-dependent $s$- and $d$-wave~\cite{Maier2015}, as well as $p$-wave~\cite{Green2020} atom-loss spectra. The maximum loss rate coefficient of the $s$- and $d$-wave Feshbach resonances (FRs) both vary with temperature, albeit in opposite manners~\cite{Maier2015}. In contrast, for the $p$-wave resonance, the maximum loss rate coefficient is independent of temperature under the condition $k_B T \gg \Gamma_{\mathrm{br}} \gg \Gamma(E)$~\cite{Green2020}. In our experiment, we did not observe a clear temperature dependence of the loss maxima for all resonances when the temperature exceeded 700~nK. However, while scanning the $m_f = 1$ resonances in the case of a dual Bose-Einstein condensate (dBEC), we needed to extend the holding time from the typical 50~ms to 120~ms to achieve the same loss ratio. This suggests a reduced maximum atom-loss rate coefficient in the quantum degeneracy regime.

Under the condition $k_B T \gg \Gamma_{\mathrm{br}} \gg \Gamma(E)$, the Lorentzian function in Eq.~(\ref{eq2}) can be approximated as a delta function peaked at $E = \mu(B - B_0)$. Substituting this into Eq.~(\ref{eq1}) and averaging the three-particle recombination rate $L_3 (E,B)$ with a Maxwell-Boltzmann distribution over the relative three-body collision energy $E$, the position of maximum atom loss correlated with temperature follows:
\begin{equation}
B = B_0 + (2+\lambda) \frac{k_B T}{\mu},
\label{eq3}
\end{equation}
where $\lambda = 1$ for the $p$-wave FRs. From the fitted slopes, we obtain the relative magnetic moment between bound trimers and entrance channels: $\mu = 3732~\text{kHz/G}$ for the $(1,1)$ resonance and $\mu = 3636~\text{kHz/G}$ for the $(1,0)$ resonance. We also note that, similar to Ref.~\cite{Green2020}, these slopes lie between $\delta \mu_b$ and $2\delta \mu_b$ (as shown in Table 1 of the main text). This possibly indicates that the bound trimer can be regarded as a superposition state of one \NaRb pair in the dimer resonant state and two \NaRb dimers in the resonant state.

From low-temperature scattering theory, for a pure BEC sample with temperature close to zero, it is predicted that no atom loss should be observed as the scattering cross-section vanishes. However, due to trap-induced confinement, atoms possess non-zero relative momentum, which explains why narrow but significant loss features are still observed.

\section{Section S2: Extracting Molecular binding energy}

Molecular binding energies, $E_b$, are extracted from the wiggle-spectroscopy measurements by fitting the atom number (or cloud size) to a model that accounts for thermal broadening. Specifically, we employ a Gaussian convoluted with a Boltzmann function~\cite{Gross2010}:
\begin{equation}
N = N_0 - A \int e^{-\frac{(h\nu - h\nu_0 \pm E)^2 }{2 \sigma^2} * \sqrt{E}e^{-\frac{E}{k_B T}}} dE,
\label{eq4}
\end{equation}
where $h\nu_0 =  |E_b|$ is the molecular binding energy corresponding to zero temperature, $N_0$ represents the atom number far from resonance, A is the modulation amplitude, $\sigma$ is the resonance width, and T is the temperature of atoms. In this model, the Gaussian component accounts for the intrinsic spectroscopic feature of molecular conversion along with broadening effects from technical noise, such as shot-to-shot atom number fluctuations and magnetic field instabilities. The Boltzmann component accounts for the asymmetric line shape inherent to finite-temperature gases.

The choice of the sign $(\pm E)$ in the Gaussian term depends on the nature of the molecular state. For bound states ($E_b < 0$), Since the relative kinetic energy of the atoms E is always positive, a larger modulation frequency $\nu$ is required to compensate for the thermal energy and drive the transition. Consequently, a minus sign is applied when fitting atom-loss spectroscopy. While for quasi-bound states ($E_b > 0$), a smaller modulation frequency $\nu$ triggers the resonance. Therefore, a positive sign is adopted when fitting the atom-size spectroscopy.

\section{Section S3: Calculation of $\delta \mu_b^{\mathrm{th}}$}

\begin{figure}[th]
\centering
\includegraphics[width=0.9\linewidth]{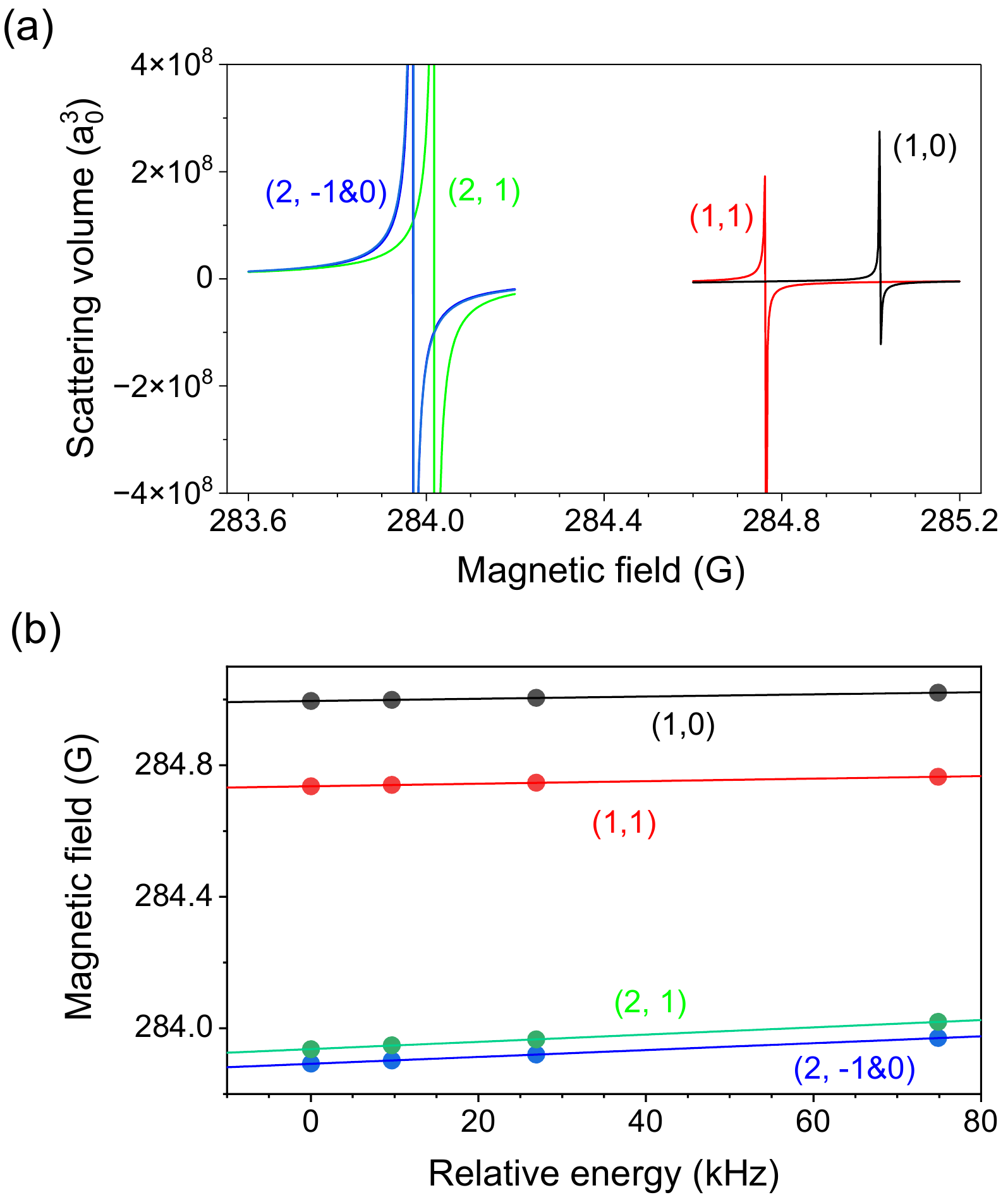}
\caption{Coupled-channel calculation of the $p$-wave scattering volume near the resonances. (a) Calculated resonance structure at a collision energy of $h\times 74.9$ kHz. (b) Resonance positions as a function of collision energy. The solid lines are linear fits, and the slopes are used to derive the inverse of the magnetic moment difference $\delta \mu_b^{\mathrm{th}}$ between the open and closed channels. The resulting $\delta \mu_b^{\mathrm{th}}$ values are summarized in Table 1 of the main text.}
\label{figs1}
\end{figure}

Theoretical molecular binding energies were calculated using the \textsc{molscat} package~\cite{Hutson2019}, with model parameters adapted from a newly calibrated molecular potential \cite{Guo2022}. By varying the relative kinetic energy between atoms in the open channel, we simulated the scattering volume as a function of the magnetic field to determine resonance locations. One example when relative kinetic energy = 74.9 kHz is shown in Fig.~\ref{figs1}(a), from which the resonance position can be fitted using a simple gaussian equation. The summarized results for $m_f = 1$ and $m_f = 2$ resonances are summarized in Fig.~\ref{figs1}(b). Linear functions are used to fit the correlation between resonance shifts and relative kinetic energies, allowing for the extraction of the theoretical magnetic moment difference $\delta \mu_b^{\mathrm{th}}$ between the open and closed channels.

We notice that in Fig.~\ref{figs1}(a), the (1,-1) resonance is missing. This is because for our entrance channel with Na and Rb both prepared in $\ket{F=1, m_F=1}$, the spin projection of the incoming atom pair is $m_f = m_{F,\text{Na}} + m_{F,\text{Rb}} = 2$. For a $p$-wave collision, the conserved total projection is $M = m_f + m_l$. Because $m_l$ is limited to $\pm 1$ and $0$, the entrance channel can only access $M = 1, 2,$ and $3$. In the presence of anisotropic spin-spin coupling, $m_f$ and $m_l$ in the closed channel are not separately conserved, but their sum is still constrained by the conserved total projection. As a result, a closed-channel state labeled predominantly by $(m_f, m_l) = (1, -1)$, which has $M=0$, cannot couple to our entrance channel. 

For the $m_f=2$ manifold, the three components $m_l = -1, 0,$ and $1$ are in principle all allowed, since they correspond to $M=1, 2,$ and $3$, respectively. However, the resonance assignment here should not be understood in terms of a simple first-order perturbative splitting of an isolated $p$-wave level. As explained in \cite{Wang2013}, due to strong mixing of nearly degenerate molecular levels from different asymptotes, this group exhibits peculiar behavior which makes the detailed structure highly sensitive to the interaction potentials, the effective spin-spin coupling, and the hyperfine interaction. As a result, the (2, -1) and (2, 0) resonances are nearly degenerate, while the (2, 1) resonance is separated away from them.

\section{Section S4: Fast magnetic field control for creating molecules}

\begin{figure}[t]
\centering
\includegraphics[width=0.9\linewidth]{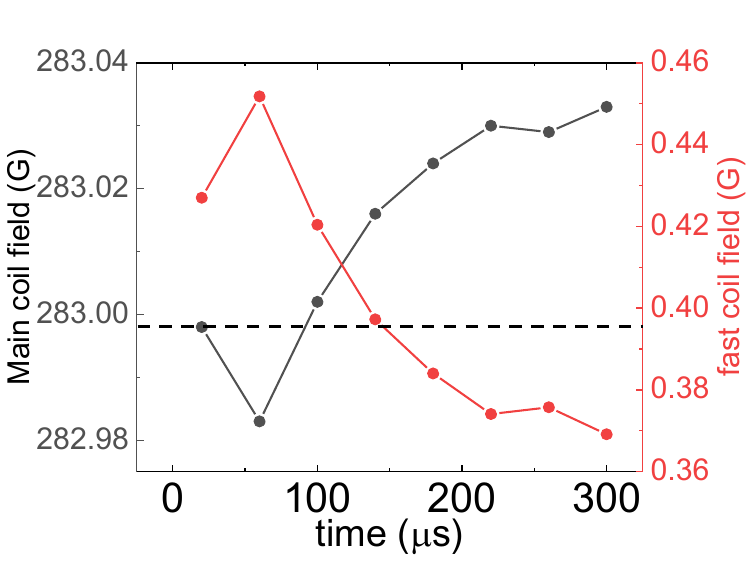}
\caption{Magnetic field control for fast atom removal. (Black) Magnetic field after the end of the main coil sweep. (Red) Magnetic field setting applied to the fast coil. (Dashed line) Net magnetic field after feed-forward compensation.}
\label{figs2}
\end{figure}

The primary magnetic field is provided by a pair of large coils. Due to limited feedback control bandwidth and Eddy current, the field typically needs over 10 ms to stabilize within 10~mG following a ramp. This residual fluctuation affects both the MA, especially in creating the pure $(2,1)$ FMs because of the small separation between the peaks, and the effective state transfer of the residual atoms to the $\ket{2,2}$ state for their rapid removal, which is necessary to increase the lifetime of the FMs. To mitigate this issue, we implemented a pair of fast-switching coils~\cite{Guo2021} and use it to actively compensate the magnetic field fluctuations using feed-forward control. An example of the magnetic field feed-forward control sequence is shown in Fig.~\ref{figs2}. This technique has allowed for the complete removal of residual atoms using microwave and blast pulses within 300~$\mu$s after the main coil sweep.

\end{document}